\documentclass{aa501}
\usepackage{graphicx}

\begin{document}

   \title{The Radio Corona of AR~Lacertae}

   \author{
         C. Trigilio\inst{1} 
          \and C.S. Buemi\inst{1}
          \and G. Umana\inst{1}
	  \and M. Rodon\`o\inst{2,}\inst{3}
	  \and P. Leto\inst{1}  
	  \and A.J. Beasley\inst{4}
	  \and I. Pagano\inst{2}
          }

   \institute{
		Istituto di Radioastronomia del C.N.R, Stazione VLBI di Noto, 
		C.P. 141 Noto, Italy 
	\and
		Osservatorio Astrofisico di Catania, Via S. Sofia 78,
		I-95123 Catania, Italy 
	\and
		Dipartimento di Fisica e Astronomia, Universit\`a di Catania,
		Via S. Sofia 78, I-95123 Catania, Italy 
	\and
		Owens Valley Radio Observatory, California Institute of
                Technology, Big Pine, 
                CA 93513, U.S.A.
      }

   \offprints{C. Trigilio}
   \mail{trigilio@ira.noto.cnr.it}

   \date{Received ; accepted }
   \abstract{
We present multifrequency VLA and VLBA observations at 8.4 GHz of the RS~CVn
system \object{AR~Lac}, that were performed in autumn 1997 simultaneously
with X-ray observations obtained from Rodon\`o et al. (\cite{Rodono}). 
Our VLBA data indicate a resolved source with dimension close to the system 
separation, while the study of the flux density curve evidences a small 
amplitude outside of the eclipse variability. 
The derived five-frequencies spectra, combined with the size information 
from VLBA data, are compared with gyrosynchrotron emission
from a two component structured source. 
A comparison  with the results of the X-ray observations allow us to exclude 
the possibility that thermal gyrosynchrotron is responsible for the radio 
emission, but it is compatible with the hypothesis of co-spatial X-ray and 
radio emitting sources. 
   }

   \maketitle
   \titlerunning{The Radio Corona of AR~Lacertae}
   \authorrunning{C. Trigilio et al.}


\section{Introduction}
The RS~CVn stars constitute one of the major classes of stellar system  
radio sources. The intense and highly variable radio emission  
is strictly related to magnetic activity and originates from the interaction of 
mildly relativistic electrons and magnetic fields (gyrosynchrotron emission)  
on one or both components. Radio observations of this kind of system can, 
therefore, allow us to directly investigate non-thermal phenomena occurring  
in stellar coronae.
\par
AR~Lacertae (G2IV+K0IV) is a totally eclipsing RS~CVn binary with a short, 
almost 2 days, orbital period (see Table~\ref{tab-par} for the parameters of 
the system). 
Its orbital characteristics and strong coronal emission 
($L_\mathrm{X}\approx 10^{31} \mathrm{erg\, s}^{-1}$) make this source an ideal 
target for the studies of coronal structure and geometry of RS~CVn close 
binaries.
\begin{table}[b]
\caption{AR~Lac parameters.}   
\label{tab-par}                               
\begin{center}
\begin {tabular}{cll}
\hline
\hline
R$_\mathrm{K}$   & 2.81 R$_\odot$    & Chambliss     (\cite{Chambliss}) \\
R$_\mathrm{G}$   & 1.54 R$_\odot$    & Chambliss     (\cite{Chambliss}) \\
Sep.             & 9.22 R$_\odot$    & Chambliss     (\cite{Chambliss}) \\
M$_\mathrm{G}$/M$_\mathrm{K}$ & 0.89 & Marino et al. (\cite{Marino})    \\
Dist             &      42 pc        & Perryman      (\cite{Perryman})  \\
P$_\mathrm{rot}$ &  $1\fd 983188$    & Marino et al. (\cite{Marino})    \\
\hline
\end{tabular}
\end {center}                    
\end{table}
The \object{AR~Lac} coronal emission has been widely studied using most of the 
X-ray satellites launched in the last twenty years. Early observations, carried 
out by Swank et al. (\cite{Swank}) using the {\it Einstein} Solid State 
Spectrometer, showed two distinct temperature distributions centred at 
$7\times 10^6$ and $1.5$ to $4\times 10^7$~K. 
On the basis of {\it Einstein} IPC observations, Walter et al. (\cite{Walter})
found that low lying coronae ($\sim 0.02\,R_\ast$) seems to be associated with 
both component stars, while a more extended corona ($\sim 1\,R_\ast$) only with 
the K subgiant. 
\par
\begin{table*}\centering	
\label{runs}
\caption{Journal of Observations}
\begin{tabular}{c|ccc|cc}
\hline
\hline
 		& \multicolumn{3}{|c|} {VLA} 			& \multicolumn{2}{c} {VLBA+VLA} \\
 		&  C Band	&  K Band	& Q Band	& X Band	& U Band 	\\
\hline
		&		&		&		&		&		\\
Nov 2-3 (U.T.)	& 21:00-08:25	& 20:30-09:25	& 20:05-09:00	& 20:05-09:25	& 21:30-08:37	\\
&  &&&\\
Nov 3-4 (U.T.)	& 21:00-08:25	& 20:30-09:25	& 20:05-09:00	& 20:05-09:25	& 21:30-08:37	\\
&  &&&\\
\hline
Number of scans                & 10     & 11    & 11    & 7         & 6         \\
Duration of a scan [m]         & 19     & 25    & 27    & 44        & 44        \\
Interval between scans [m]     & 80     & 80    & 80    & 60        & 60        \\
Duty cycle (cal-sour-cal) [m]  & 2-15-2 & 3-8-3 & 3-5-3 & 1.5-3-1.5 & 1.5-3-1.5 \\
Total Time on source [h]       & 2.42   & 2.54  & 2.4   & 2.75      & 2.5       \\
\hline
\end{tabular}
\end{table*}
EXOSAT observations by White et al. (\cite{White}) confirmed the existence
of two different coronal regions associated with different temperature 
distributions. 
The observations showed  evidence of modulations at
low energy ($<$ 1 keV), which are not present at high energy ($>$1 keV). 
On the other hand, Ottmann et al. (\cite{Ottmann}) observed evidence of the 
primary minimum in all energy bandpasses of the ROSAT PSPC.
\par
Despite the numerous X-ray observations only a few have been carried out at 
radio wavelengths.
Owen \& Spangler (\cite{Owen}) first studied the spatial structure of the 
radio corona of \object{AR~Lac}, on the basis of flux curve considerations, 
through observations performed at 4885 MHz with 5 VLA antennas. 
They did not found evidence of eclipses of either component. 
The same results were found by Doiron \& Mutel (\cite{Doiron})
on the basis of VLA observations carried out at 1.48 and 4.9 GHz, using 
all the 27 antennas of the array in B configuration.
These results suggest that the size of the emitting region is larger than 
the whole system.
\par
During a VLA 3-frequencies monitoring of \object{AR~Lac}, extended over an 
orbital period, Walter et al. (\cite{Walter87}) detected the decay of a 
relative large flare. The event was not observed at 20 cm and the radio 
spectra seems to turn over between 2 and 6 cm. 
\par  
In the present paper we present the results of a multifrequency observing
campaign on the binary system AR~Lacertae (HD~210334), 
simultaneously 
performed with the Very Large Array (VLA) and the Very Long Baseline Array 
(VLBA) 
	\footnote{
	The Very Large Array and the Very Long Baseline Array are facilities of 
	the National Radio Astronomy Observatory which is operated by 
	Associated Universities, Inc. under cooperative agreement with the 
	National Science Foundation}
radio interferometers in the autumn 1997. The observations cover
two orbital periods and 
took place contemporaneously with X ray observations 
of the same system carried out with the SAX satellite by Rodon\`o 
et al. (\cite{Rodono}).
\par

\section{Observations and data reduction}
\begin{figure*} 
\centering
 \includegraphics[width=17cm]{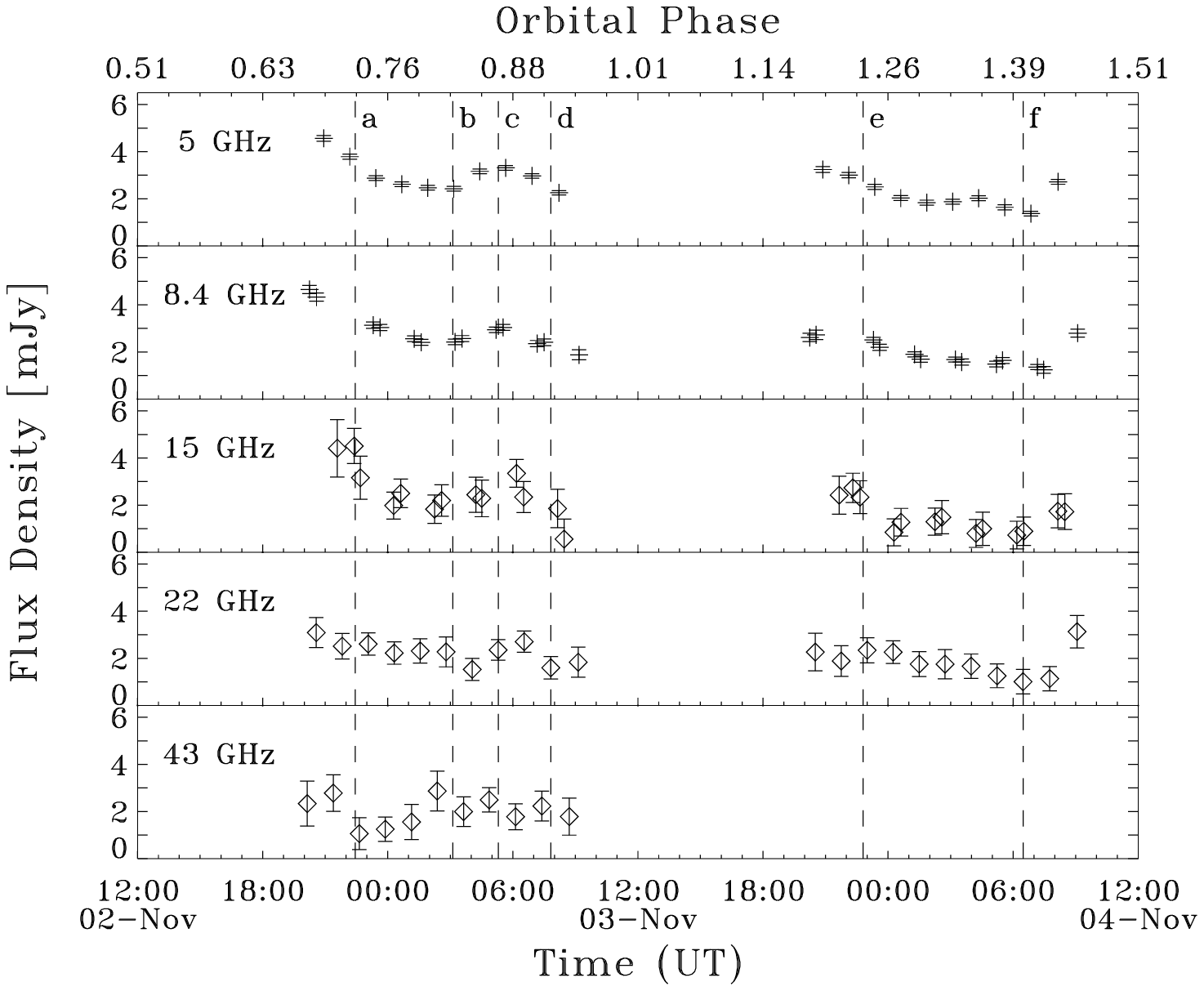}
 \includegraphics[width=17cm]{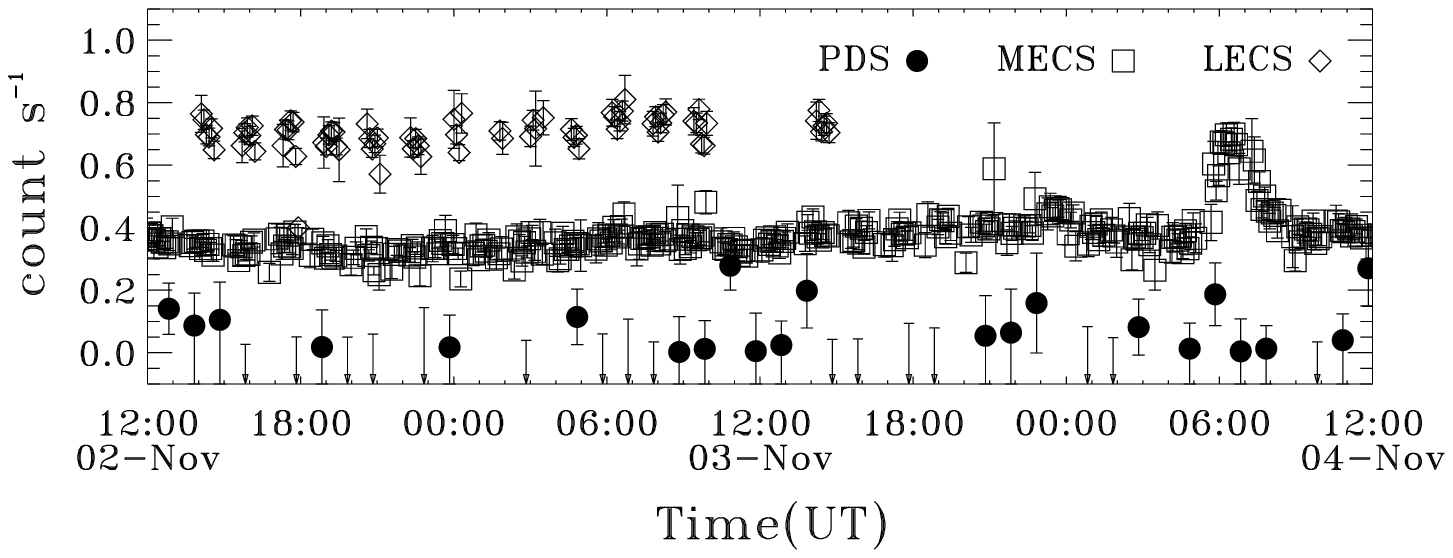}
\caption[]{Top panel: VLA flux curves of \object{AR~Lac} binned over 20 
minutes. Dot-dashed vertical lines indicate the times of the spectra labelled 
by the letters on the top and shown in Fig.~\ref{nomod}. 
Lower panel: The contemporaneous SAX light curves 
(from Rodon\`o et al. \cite{Rodono})}
\label{film1}
\end{figure*}
%
The observations were performed in 1997 in four sessions, starting each day at 
20:00 UT and ending at 9:40 UT of the following day, from Oct 31 to Nov 4. 

The aim of these observations was to obtain simultaneously radio spectra at 
five frequencies and high resolution maps of \object{AR~Lac} at 4~cm and 2~cm. 
The spectra were obtained with the VLA in D configuration 
at 5~GHz (6~cm, C band), 8.4~GHz (4~cm, X band), 15~GHz (2~cm, U band), 
22~GHz (1.3~cm, K band) and 43~GHz (0.7~cm, Q band), 
while the high resolution VLBA observations were performed
at 8.4 and 15~GHz. In order to reach the best compromise between 
high sensitivity for the VLBA array
and the best temporal coverage for the radio spectrum, we split the VLA
into two subarrays, one observing in standard interferometric mode and
alternating 3 frequencies, the other in phased array mode as element of the VLBA.

\subsection{VLA data}
The first VLA subarray, which consisted of the 13 telescopes equipped with the 43~GHz 
receivers, was used to observe alternately at 43, 22 and 5~GHz, with two
independent 50~MHz wide contiguous bands, in Right and Left circular polarization.

As phase calibrator we used \object{BL~Lac}, while 
as primary flux calibrator we used \object{3C286}, which was observed at the begin of 
each session at all the 3 frequencies.

To minimize the effects of the atmosphere on the phase stability at higher frequency 
we observed the phase calibrator more frequently at Q and K bands 
than at the other frequencies. Other details of the observational
strategy are given in Table~\ref{runs}.

Calibration and data editing were performed using the standard procedures 
of the AIPS package.

Since we intend to study only our radio observations taken 
simultaneously 
with the X-ray observations, which started on 1997 Nov 2,
we restrict our analysis to the last two sessions, i.e. Nov 2-3 and Nov 3-4. 
In the  first two days the source was found in an active period, with flux
densities reaching about 40~mJy at 6~cm. This indicates an intrinsic high 
variability of the radio emission, which will be analyzed in a following paper.

\subsection{VLBA data}

The second subarray, consisting of all the remaining 14 antennas, was used to 
observe in phased-array mode at X and U band as an element of the VLBA array.
The presence of half the phased VLA, corresponding to a 90~m telescope, 
increased the sensitivity of the VLBA.

The data from the VLA phased-array can be also analyzed as a standard 
interferometer, therefore total fluxes were computed for all the observed 
sources. The VLBA observations were performed alternating 
between X and U bands, with a 
typical 
scan lasting 45 minutes.
For each observing frequency, we observed in dual polarization mode, with
a total bandwidth of 64~MHz. Since \object{AR~Lac} has generally a low flux 
density (from few to some tens of mJy), we used the phase-reference technique 
(Beasley \& Conway, \cite{Beasley}), consisting of the rapid switching between
a strong calibrator, close to the target source, and the source itself.
In this way, in the successive analysis, phase calibration of the calibrator
can be applied to the target source.
We used \object{BL~Lac} ($3.6$ degrees apart from \object{AR~Lac}) as
reference source. 
The data tapes were correlated at the VLBA correlator at the Array Operation 
Center in Socorro (New Mexico).

As for VLA data, calibration and data editing were performed using the standard 
procedures of the AIPS package. 
The flux calibration was performed by using the measured system temperature and 
gain curves for VLBA telescopes and from the measurements of the ratio 
$T_\mathrm{ant}/T_\mathrm{sys}$ made on the calibrator \object{BL~Lac} for the 
phased-array, once the flux density of this source was determined from the VLA data. 
Delay and delay rate were determined for \object{BL~Lac}, which was
then self-calibrated. The final map of the self-calibration was used as
a model for the final global fringe fitting, whose solutions were applyed
to our target. A preliminary map of \object{AR~Lac} at 8.4~GHz was made to 
check its position, and it was found about 14~milliarcsec (mas) away from the 
phase tracking center.
This was due to the fact that the coordinates used at the correlator were the 
heliocentric ones, and the displacement of the source was due to the annual parallax. 
The visibilities of \object{AR~Lac} were then phase-rotated in order to bring 
it to the phase center. The operation makes it possible 
to time average the data over a relatively long time (tens of minutes) without
lose signal in the longest baselines, allowing us to analyze the amplitude
of the visibility as a function of the baseline length. 
If the source is not at the phase center a rapid 
change of phase would occurr 
at the longest baseline.

The source was successfully detected at 8.4~GHz, but not at 15~GHz,
probably 
because of a poor phase stability 
at this frequency, which we could not correct with 
the phase referencing technique.
%
%
\begin{figure*} 
\centering
 \includegraphics[width=17cm]{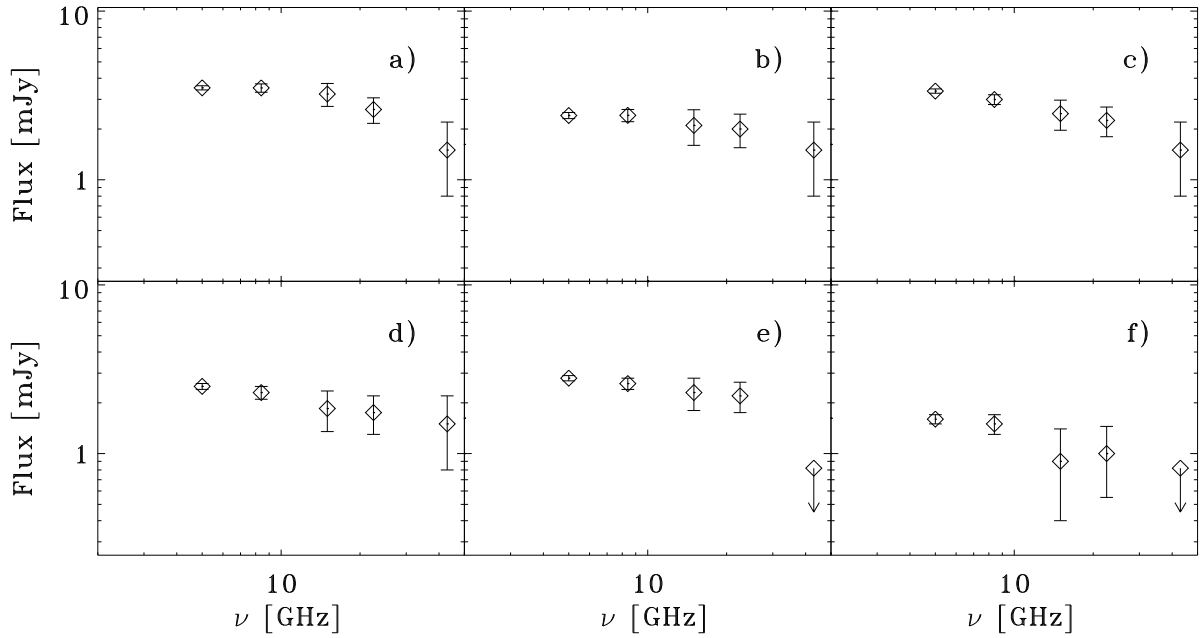}
\caption[]{\object{AR~Lac} VLA spectra obtained at times indicated by dashed 
vertical lines in Fig.~\ref{film1}} 
\label{nomod}
\end{figure*}
%

\section{Results}
\subsection{Total flux and Spectra from VLA data}
In order to investigate the temporal radio flux behaviour of the source
and locate possible rapid flux variations, we analysed each scan with the 
AIPS program DFTPL. This procedure performs the direct Fourier transform of the 
visibility function in a limited portion of the UV plane. The results of such 
analysis led to the conclusion that, even if the radio flux of \object{AR~Lac} 
is variable, it does not change significantly on timescale shorter than 45 
minutes, which is about the time length of each scan.
\par 
In Fig.~\ref{film1} the flux curves, obtained by averaged flux in 20 min bins
and using the new ephemeris by Marino et al. (\cite{Marino}), are shown.
\par
The source was detected at all frequencies during both sessions, 
except during the last session in the Q-band, where only an upper limit of 
0.6~mJy for the flux density can be given. This value corresponds 
to the rms of the cleaned map integrated over the entire 14 hours. 
A clear flux variation, more evident at the lowest
frequencies (up to a factor of 2 in C-band), seems 
to indicate the presence of inhomogeneous features on one or 
both of the system's components. 
Unfortunately, AR Lac was not visible from the VLA during eclipses.
We observed a clear flux maximum at phase $\approx$0.88, and a clear decay at 
the beginning of each of the two sessions, suggesting a flux modulation versus
phase. 
No significant circular polarization at any observed frequency was detected.
\par
Since each band was not observed continuously, it is not possible
to build radio spectra with simultaneous flux measurements at all the 
frequencies. However the variations of the flux are slow, and it was possible
to make reasonable interpolations in order to obtain the radio spectra.
Six spectra corresponding to
different times are shown in Fig.~\ref{nomod}. They all show negative
spectral indexes $\alpha$ (S $\propto\, \nu^{\alpha}$), 
implying that the source is optically thin in the available 
range of frequencies
($\alpha$ varies from $-0.16$ to $-0.45$). 
There is not evidence of a turn-over frequency,
which is presumable lower than 5 GHz. 
The spectra show the approximately same shape, and the main difference
is flux density variation.

\subsection{Spatial structure from VLBA data}
\begin{figure*}\centering \leavevmode
\resizebox{17cm}{!}{
	\includegraphics{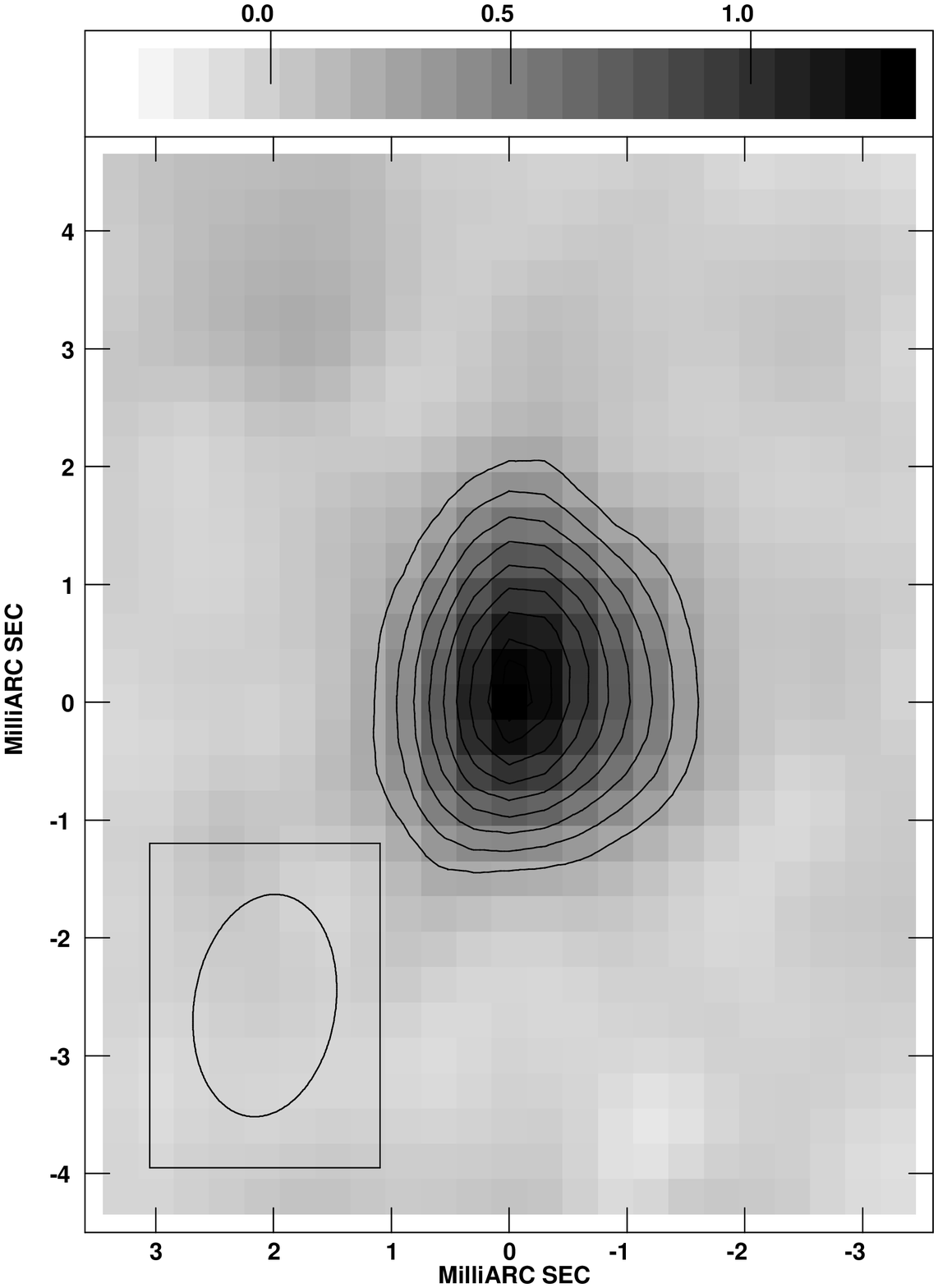}
	\includegraphics{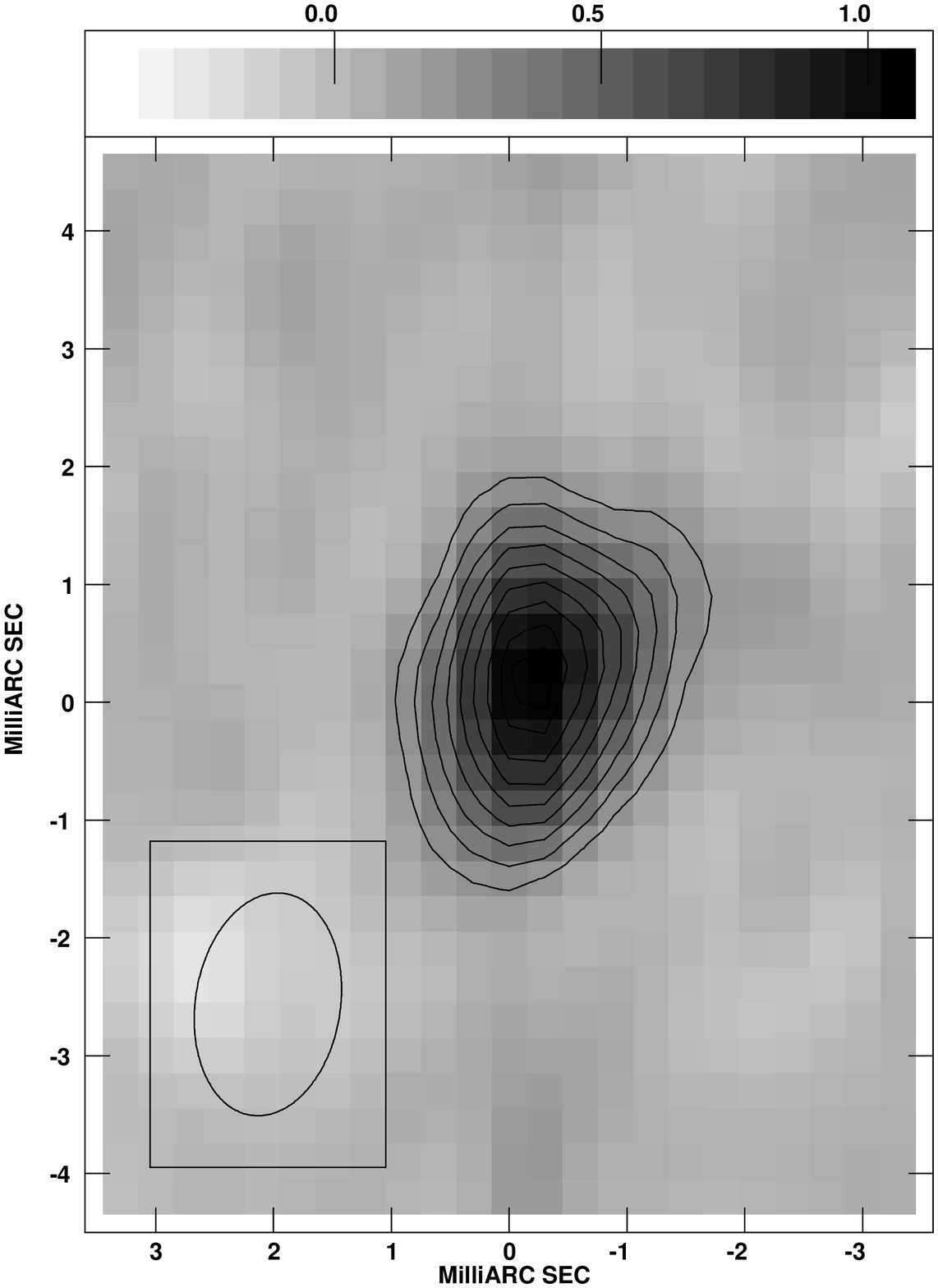}
} 
 \caption[ ]{
VLBA maps of \object{AR~Lac} at 8.4~GHz for Nov 2-3 (left) and Nov 3-4 (right). 
The r.m.s. of both maps is 0.06 mJy/beam and the peak intensity are 1.26 and 
1.07 mJy, respectively. For the first map the contours correspond to 
0.25, 0.38, 0.50, 0.63, 0.76, 0.88, 1. 01, 1.13, 1.21 mJy, 
for the second map correspond to 
0, 0.21, 0.32, 0.43, 0.54, 0.64, 0.75, 0.86, 0.96, 1.03 mJy. 
On the lower left corner the restoring beam is shown
}
\label{map}
\end{figure*}
\begin{figure*} 
\centering \leavevmode
\resizebox{17cm}{!}{
	\includegraphics{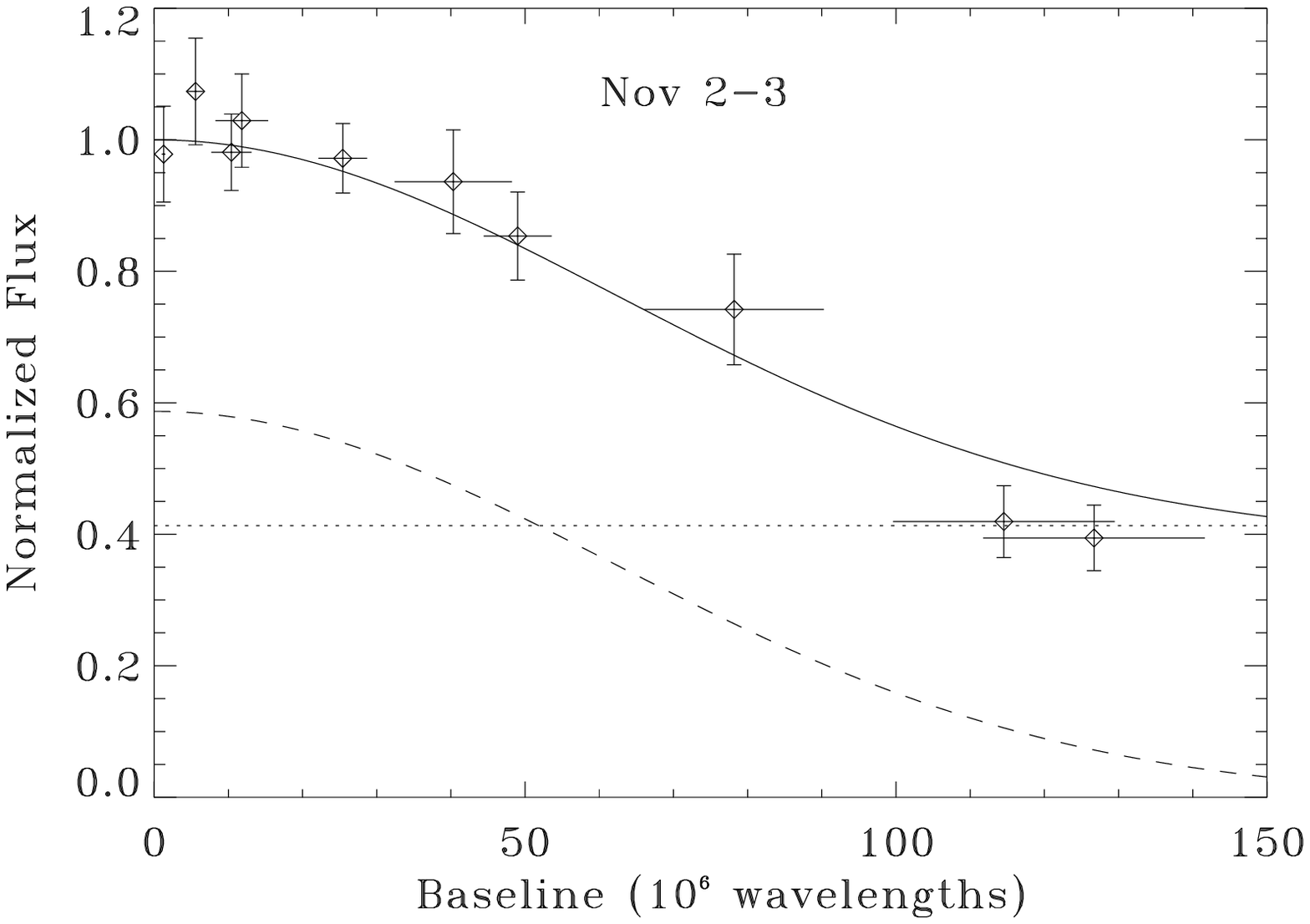}
	\includegraphics{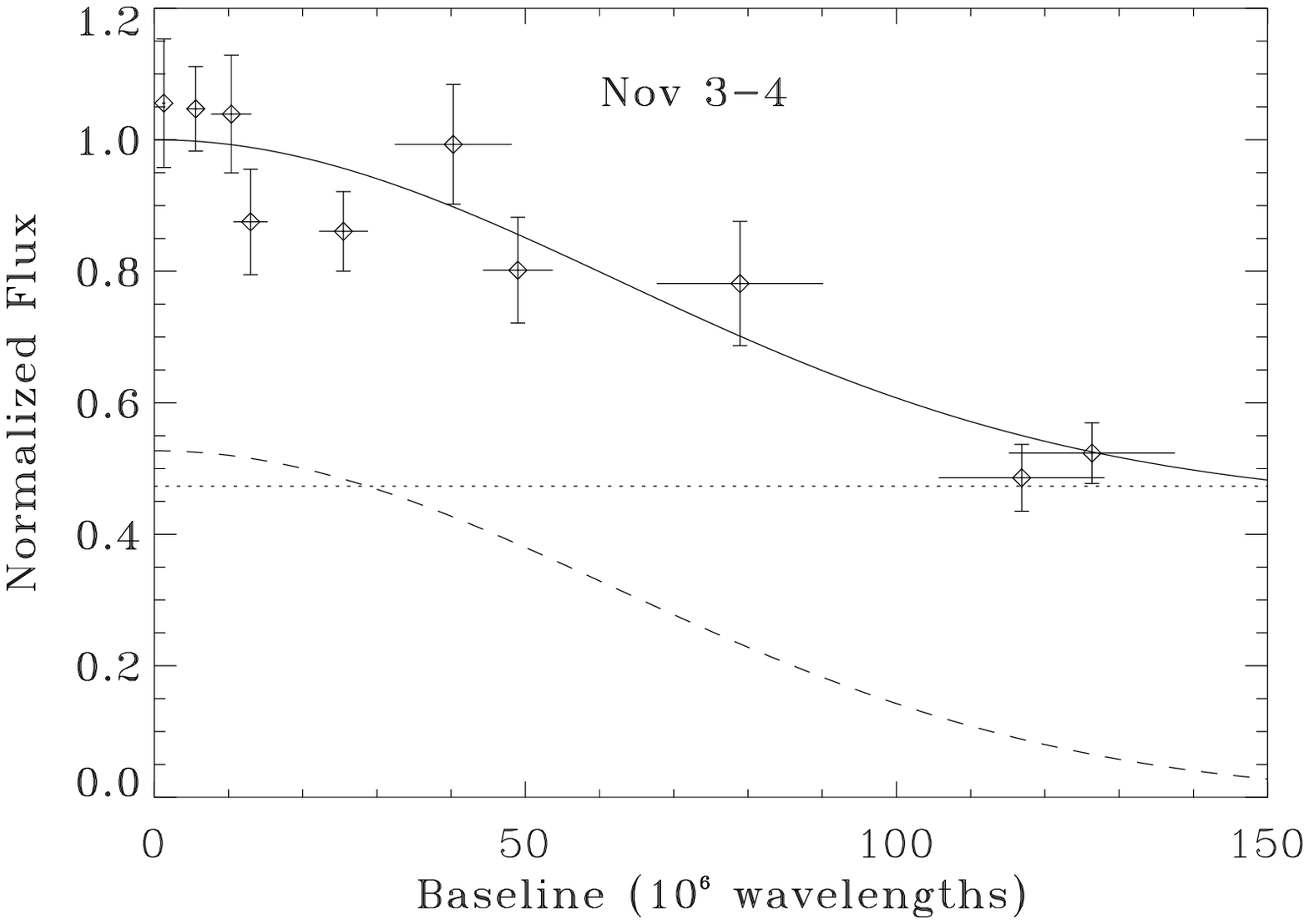}
} 
 \caption[ ]{
Correlated flux at 3.6 cm on the VLA-VLBA baselines as function of $u-v$ 
distance, for Nov 2-3 and Nov 3-4. 
Tle lines represent the visibility of the core-halo model (continuous lines),
of the single core (dot lines) and halo (dashed lines) as discussed in 
section \ref{discussion}
}
\label{f-uv}
\end{figure*}
In Fig.~\ref{map} the VLBA cleaned map of \object{AR~Lac} for the two sessions, 
which were obtained from the phase reference calibration, are shown. 
The FWHMs of the beam for both the maps are 
$1.9 \times 1.2$~mas, with a position angle of $-8\fdg 78$ and 
$\sigma\approx\,0.06$~mJy. We did not measure significant circular
polarisation from the VLBA data.
The source appears resolved and the two maps do not show significant changes 
in their structure.
\par 
The evidence of a resolved structure is confirmed by the analysis of the visibility 
amplitude versus the projected baseline spacing for the VLBA 
baselines involving VLA, which is shown in Fig.~\ref{f-uv}. 
The correlated flux has been first coherently averaged through the scan and 
then the amplitudes normalized to the zero spacing value, which was 
assumed to be equal to the total flux obtained from the VLA measurements in 
the same scan. 
Then, the visibility amplitudes have been averaged for each baseline,
whose length variation during the observation is indicated by the horizontal
bars.
In order to measure the size of the emitting regions, or to derive the upper
limit, we fitted the data with a two dimensional Gaussian models by using the 
AIPS routines UVFIT and JMFIT. 
The first routine gives an acceptable fit of UV-data with an elliptical 
Gaussian with 
FWHM of $1.4 \times 1.2\pm 0.2$~mas, while the second one
fits the source image with a Gaussian component having FWHM of
$1.3 \times 1.1 \pm 0.2$~mas. The two models are then consistent
each other and with a circular symmetric source. 
Using a circular Gaussian to fit the data, we obtain a source size of 
$1.2 \pm 0.2$~mas, 
corresponding to $4\,R_\mathrm{K}$ (see Table~\ref{tab-par}) or $1.2$ times 
the system separation ($\simeq\,1$~mas).
Our data are also compatible with two component models, where the
first component is resolved, with dimension between 1.3 and 1.6~mas,
while the second component is a point source. On the basis of $\chi^2$ results
we cannot exclude the presence of such compact core.

\begin {table}
\caption[]{Parameters of the core-halo
}
\label{fit}                               
\begin{center}
\begin {tabular}{c|cc}
\hline
\hline
				&  CORE &   HALO \\
\hline
				&	& 	 \\
$B$ [Gauss]		 	& 200 	&   30   \\
$N_\mathrm{rel}$ [cm$^{-3}$] 	& $8\times 10^5$ 	&  $3\times 10^4$  \\
				&	& 	 \\
\hline
Spectrum 			& d 	&  d     \\
        			& [mas] & [mas]  \\
\hline
$a$ 				& 0.16 	& 1.4    \\  
$b$ 				& 0.14 	& 1.1    \\  
$c$ 				& 0.15 	& 1.3    \\  
$d$ 				& 0.14 	& 1.2    \\  
$e$ 				& 0.16 	& 1.3    \\
$f$ 				& 0.13 	& 1.1    \\
				&	& 	 \\
\hline
				&	& 	 \\
$<R>/R_\odot$		& 0.7	& 5.4	 \\
$EM$ [cm$^{-3}$]$^{\dag}$ 	& $2.3\times10^{53}$ & $3.1\times10^{53}$  \\
$T$ [K]$^{\dag}$  & $(0.93-1.02)\times10^{7}$  &$(2.47-2.73)\times10^{7}$ \\
$N_\mathrm{e}$ [cm$^{-3}$]	& $2.3\times10^{10}$ & $1.2\times10^{9}$  \\
				&	& 	 \\
\hline
\end{tabular}
\begin{list}{}{}
\item[$^{\dag}$] Rodon\`o et al. (\cite{Rodono})
\end{list}
\end {center}                    
\end{table}

\section{Discussion}
\label{discussion}
%
%
\begin{figure*} 
\centering
 \includegraphics[width=17cm]{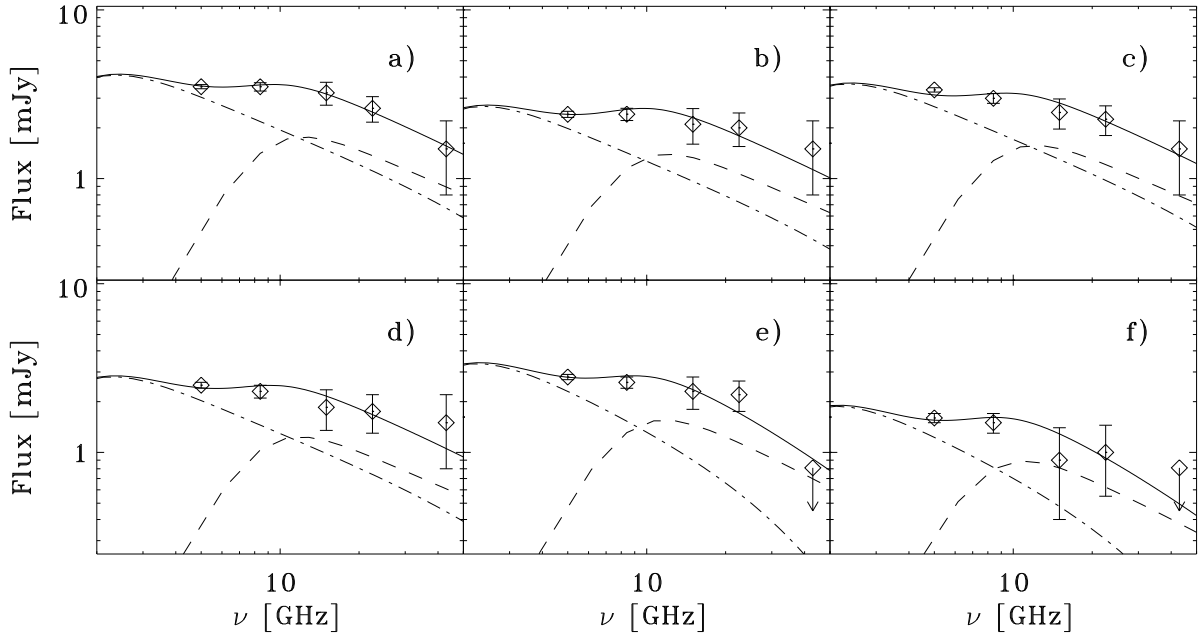} 
 \caption[ ]{Comparison between the observed radio spectra of \object{AR~Lac}
shown in Fig.~\ref{nomod}
and the computed spectra obtained by assuming a core-halo structure for the
radio source (thick line). The contribution of the halo (dot-dashed line)
and core (dashed line) to the composite spectrum are also shown}
\label{spec}
\end{figure*}
\begin{figure}
\centering 
\resizebox{\hsize}{!}{\includegraphics{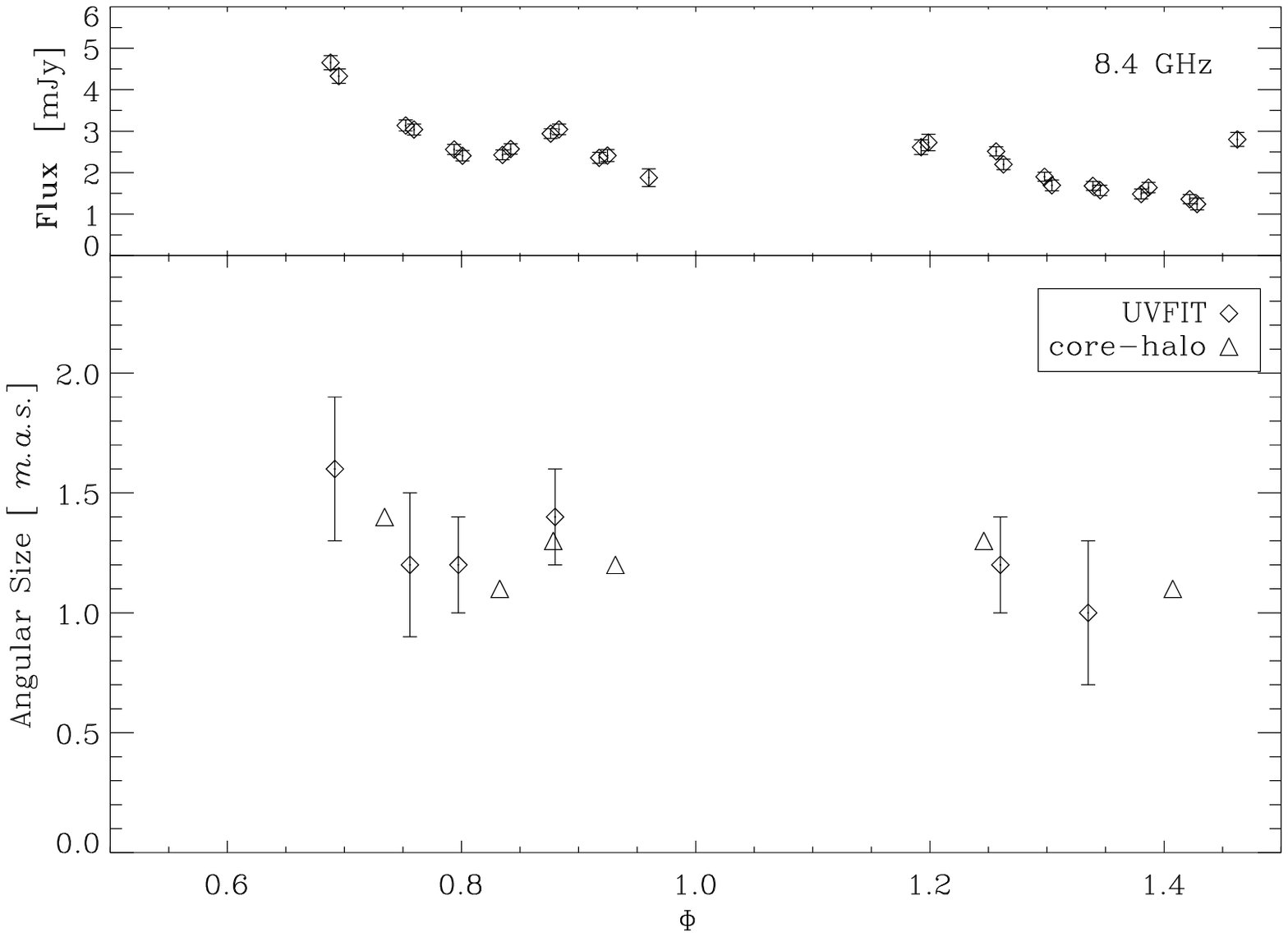}} 
 \caption[ ]{Lower panel: Source size estimates as a function of orbital phase. 
The diamonds represent the value obtained from UVFIT, the triangles represent 
the value of source size we used to fit the observed spectra with the core-halo 
model. 
Upper panel: The corresponding total flux density at 8.4~GHz
}
\label{dim}
\end{figure}

%
The availability of both microwave spectra and spatial information 
at the same time gives the opportunity to model the coronae of \object{AR~Lac}. 
As shown in previous papers (Umana et al. \cite{Umana93}, Umana et al. 
\cite{Umana99}), the observed flat radio spectra of Algols and RS~CVn type 
binary systems cannot be reproduced by an homogeneous source model. 
VLBI observations of the close binary systems Algol and UX~Arietis pointed out 
the existence of a two component structure in the coronal layers (Mutel et al. 
\cite{Mutel}; Lestrade et al. \cite{Lestrade}): 
a compact core, coinciding with the active star, and a larger halo, having 
approximatively the size of the entire system.
\par
We used the core-halo model developed by Umana et al. (\cite{Umana93}) to fit 
the observed spectra of \object{AR~Lac}, in order to check if the flux 
variability can be attributed to the variation of one of the 
parameters.
Although it is not possible to derive an unique solution, we have limited the 
sizes of core and halo on the basis of the VLBA observations results. 
In particular, for the halo we assumed the size measured from the VLBA data, 
and for the core a size smaller than the resolution limit of the 
interferometer.
\par 
For the magnetic field strength we assumed
$B\sim B_\mathrm{phot}(\frac{R}{R_\ast})^{-3}$, where 
$B_\mathrm{phot}\sim 600-1000$~Gauss, as derived 
for other RS~CVn systems (Gondoin et al. \cite{Gondoin}, Donati et al. 
\cite{Donati92}). 
\par
According to these constraints, we derived the best core-halo model fit
labeled {\it a} in Fig. 2.Then 
we used the derived values of the average magnetic field strength $B$
and energetic electron number density ($N_\mathrm{rel}$), 
in fitting the other 5 spectra, 
under the hypothesis that these physical properties of 
the coronal emitting regions are stationary. 
Then we tried to fit the other spectra by varying only the structure size.
The results of our analysis are summarised in Table~\ref{fit} and shown in
Fig.~\ref{spec}. For the spectra from {\it a} to {\it d}, we can get a good 
agreement between the observed data
and a core-halo structure by assuming that the flux variations are due to 
structure size changes in the range of 1.1 and 1.4 times the stellar
radius, corresponding a variation of 0.3~mas, which is below
the errors of our measurements. 

The low s/n ratio of the VLBA data does not permit  
to investigate the behaviour of the emitting region size for each single scan,
and so as a function of the orbital phase.
The source size is very close to the beam width, so it was not 
possible to obtain an estimate of the source size for the time ranges 
during which the flux density was lower then $\approx 2$~mJy. 
Nevertheless, the results obtained 
for the scans with sufficient s/n ratio, that are shown in Fig.~\ref{dim}
by diamond symbols,
suggest that the source size remains almost constant, as the measured
changes fall within errors.

On the other hand, it should be noted that the
 slow modulation of the modelled halo size is in good agreement with 
 the results obtained from an independent analysis of 
 the VLBA data and plotted in Fig.~\ref{dim}. 
To fit the spectra {\it e} and {\it h}, we had to assume an ''ageing''
of the relativistic electron population and a variation of its number
density from $8.0\times 10^{5}$ to $7.0\times 10^{5}$~cm$^{-3}$, that is
needed to explain the faster decay at the higher frequency.
\par
If the core-halo model is suitable to account for the radio corona of
\object{AR~Lac}, the visibilities of the VLBA data should fit by a
two gaussian model corresponding to the core and halo.
For the Nov 2-3 data, the average size of the core from the analysis of
the spectra {\em a, b, c, d} is 0.15~mas, and the average flux density
at 8.4~GHz is 1.21~mJy; for the halo, 1.25~mas and 1.74~mJy.
For the Nov 3-4 data, the same parameters from the analysis of the spectra
{\em e} and {\em f} are 0.145~mas, 1.05~mJy for the core and 1.20~mas,
1.17~mJy for the halo. 
We then model the  normalized visibility function at 8.4~GHz with that 
corresponding to 
the core-halo model (Fig~\ref{f-uv}). It is evident
that the VLBA data are consistent with the core-halo scenario 
derived from  the analysis of the radio spectra.
\par
VLBA data indicate a source size close to the  
separation of the binary components, 
suggesting the possibility of an emitting region located between 
the system components. 
UV emission from plasma close to the Lagrangian point in between the system
components 
{\bf has been suggested} 
for \object{AR~Lac} (Pagano et al.
\cite{pagano}) and other RS~CVn-type systems (Bus\`a et al. \cite{busa}).
\par
In partial overlap to our observations, X-ray observations of \object{AR~Lac}
were performed with the Beppo SAX satellite (Rodon\`o et al. \cite{Rodono}). 
This gave us the opportunity to 
determine whether 
the physical parameters of the 
radio emitting regions, derived from the comparisons between the observations 
and the core-halo models, are consistent with a co-spatial model for both the 
X-ray and radio emitting source.
\par
Spectral analyses performed by several authors (Swank et al. \cite{Swank}, 
Singh et al. \cite{Singh}) 
showed that the X-ray emission from close binary systems requires
at least two plasma components characterised by different temperature and
volumetric emission measure ($EM$) to be modelled. 
On the basis of the first observation run, started on Nov 2 at 06:07 and ended 
on Nov 4 at 17:50, Rodon\`o et al. (\cite{Rodono}) derived for the first 
component 
$T_1=(0.93-1.02)\times 10^7$~K and $EM_1=2.3\times 10^{53}$~cm$^{-3}$ 
and for the second  
$T_2=(2.47-2.73)\times 10^7$~K and $EM_2=3.1\times 10^{53}$ cm$^{-3}$.
\par
Assuming that the higher temperature component is associated with the halo and 
the the lower temperature component with the core, we can check if the magnetic 
field, as derived from the radio data, is strong enough to contain the
X-ray source. This means that $\beta$, i.e. the ratio between the density
of kinetic energy ($P_\mathrm{c}\approx 2N_\mathrm{e}kT$) to the density of 
magnetic energy ($P_\mathrm{M} \approx \frac{B^2}{8 \pi}$) has to be less then 
unity. 
If the plasma density $N_\mathrm{e}$ is constant over the emitting volume
$V$, $EM=V\times N_\mathrm{e}^2$, and assuming the size (radius) from the
analysis of the radio data of 
$4.7\times 10^{10}$ and $3.8\times 10^{11}$~cm 
for the core and the halo respectively, we get 
$N_\mathrm{e}\approx 2.3\times 10^{10}$ and 
$\approx 1.2\times 10^{9}\mathrm{cm}^{-3}$
(see Table~\ref{fit}).
\par
We obtain $\beta = 0.04$ for the core and $\beta =0.23$ for the halo. The
physical parameters obtained from our analysis are therefore consistent with
the hypothesis of co-spatial X-ray and radio source.

\par
\begin{figure}\centering 
\resizebox{\hsize}{!}{\includegraphics{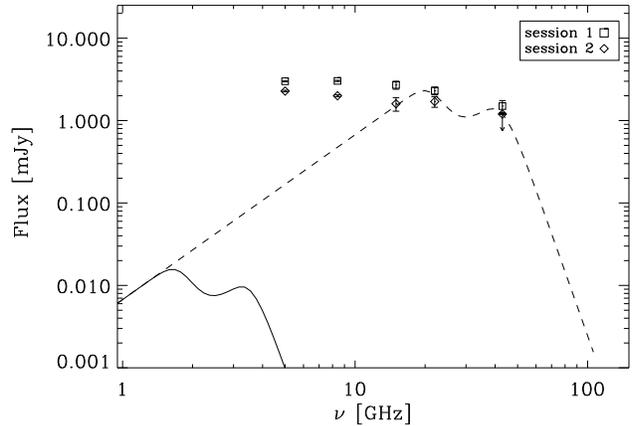}} 
 \caption[ ]{Simulated spectra from a core-halo structure for thermal 
gyrosynchrotron obtained with a magnetic field of 600 Gauss in the core and
200 Gauss in the halo (thick line) and $10\,000$ Gauss in the core and
300 Gauss (dashed line) for the halo. 
Overlaid are the average spectra for the two sessions}
\label{term}
\end{figure}

We will furthermore test  the possibility that the radio emission can be 
attributed to the same thermal electron population responsible for the observed 
X-ray emission. 
The brightness temperature of the resolved radio source at 3.6 cm, obtained 
from the relation 
$$
T_\mathrm{B}=1.97\times 10^6 
\frac{F_\mathrm{mJy}\lambda^2_\mathrm{cm}}{\theta_\mathrm{mas}^2}
$$
is  T$_\mathrm{B}\, \simeq \, 5.39\times 10^7$~K 
and T$_\mathrm{B}\, \simeq \, 3.70\times 10^7$~K 
for the first and the second session, respectively. These values are higher of
a factor of two respect to the temperature derived from the Beppo SAX data,
but the order of magnitude is the same. 
We simulated 
the spectra
of a core-halo structure by adopting thermal gyrosynchrotron emission and 
using the expression given by Dulk (\cite{Dulk}) for the emission and 
absorption coefficients. By combining the physical parameters
derived from Rodon\`o et al. (\cite{Rodono}) for the thermal coronal plasma 
with our VLBA results we can fix 
the temperature, the thermal plasma
density and the maximum size of the emitting regions. The only free
parameter left is, thus, the magnetic field strength. The spectrum obtained 
for B=600 Gauss for the core and B=200 Gauss for the halo is plotted
in Fig.~\ref{term} and compared with the spectra obtained from the 
VLA maps over the two sessions. 
\par
It is evident that gyrosynchrotron emission from the same thermal population 
responsible for the X-ray emission is not able to account for the observed 
spectra, unless magnetic field strength higher then 1000 Gauss are considered.
Moreover, even assuming such an intense magnetic field, it is not possible to 
reproduce the quite flat observed spectra.
Beasley \& Guedel (\cite{beasley2000}) reached a similar result from 
simultaneous radio and X-ray observations of the RS~CVn-type
binary system \object{UX~Ari} during quiescence.

\section{Conclusions}
In this paper we presented the results of VLA multiwavelength observations and
the first 8.4 GHz VLBA images of \object{AR~Lac} we obtained in November 1997. 
These 
images show spatially resolved structure with a diameter of $1.2\pm 0.2$~mas, 
as derived from model fits. Since the source was very compact,
with dimension close to the beam size, and the flux density level was very low,
it was not viable 
to derive the source diameter for each phase. Still, whenever possible,  
the source size was derived (within 1--2 $\sigma$).
No significant variations of the source size are apparent.\\
The flux density curves show a slight flux modulation, more evident at lower 
frequencies, which suggests the presence of inhomogeneous structures. The
five-frequency spectra show a slightly negative spectral 
index, which is
characteristic of an optically thin source. 

By combining the spectral information the estimate of the source size 
and the results obtained from simultaneous X-ray observations 
(Rodon\`o et al. \cite{Rodono}), 
we conclude that:
\begin{itemize}
\item 
the spectral and spatial information of the radio corona of \object{AR~Lac} 
indicate a structured morphology, which can be modelled with a core-halo 
source;
\item   
the physical parameters, as derived from the fit of the observed
spectra with the core-halo model, are consistent with the hypothesis of
co-spatial X-ray and radio source; 
\item 
the observed radio emission cannot be attributed to the same thermal electron 
population responsible for the observed X-ray emission.
\end{itemize}
The application of the core-halo model to the observed 
spectra allows us 
to reproduce the flux modulation observed between phases 0.7 and 0.9,
by keeping fixed
the physical parameters which characterise the two structures but changing
just the size within the range defined from the VLBA resolution. 
Further observations will be able to verify the stability of the structure
that can originate the modulation.
In particular, since the numerous X-ray observations of this binary system 
seem to indicate structural changes of its corona
(see Rodon\`o et al. \cite{Rodono} for a summary) it will be extremely 
important to find out if this is true also for the radio corona and to check 
if our hypothesis of co-spatial emitting region is confirmed. 

In addition, multiwavelength VLA observations to be carried out during  
eclipses would  be a crucial test the hypothesis of an active region
located in between the system components.
\par

\end{document}